\documentclass[12]{article}
\usepackage{graphics,graphicx,amsbsy,amssymb,color}

\newcommand{\JJ}{{\boldmath \mbox{$J$}}}

\newcommand{\uu}{{\boldmath \mbox{$u$}}}

\newcommand{\rr}{{\boldmath \mbox{$r$}}}

\newcommand{\cc}{{\boldmath \mbox{$c$}}}

\newlength{\defbaselineskip}
\newcommand{\setlinespacing}[1]%
           {\setlength{\baselineskip}{#1 \defbaselineskip}}

\title{\textbf{Rheological modeling with GENERIC and with the Onsager principle}}
\author{Miroslav Grmela \footnote{ e-mail:
miroslav.grmela@polymtl.ca}\\
\'{E}cole Polytechnique  Montr\'{e}al,
  C.P.6079 suc. Centre-ville,\\
 Montr\'{e}al, H3C 3A7,  Qu\'{e}bec, Canada}

 \date{}

\begin{document}

\maketitle

\begin{abstract}

In this paper we compare three frameworks for modeling flows of complex fluids: (i) local conservations of mass, momentum and energy, (ii) GENERIC, and (iii) Onsager principle.  The first is based on the mass, momentum, and energy conservation implied by mechanics,
the second on the observed approach  (in the state space)  of externally unforced fluids to thermodynamic equilibrium states, and the third on the approach (in the space of vector fields) of externally driven fluids to dynamics involving less details.
The comparison is illustrated on  isothermal  polymeric fluids.

\end{abstract}

\section{Introduction}

How do complex fluids (like for instance polymeric fluids and suspensions) flow? This question gave birth to rheology. The same question but with simple fluids (like for instance water) replacing the complex fluids gave birth to hydrodynamics \cite{Euler}. Rheology is  an extension of hydrodynamics from simple fluids to complex fluids. The mathematical models of fluids that are emerging in hydrodynamics and in rheology are  expected to reproduce first of all their  observed flow behavior. There are  however also other types of the behavior of  fluids that can be readily  observed.
For instance it is the overall mass, momentum, and energy conservation or the approach of externally unforced fluids to thermodynamic equilibrium at which their observed behavior is found to be well described by the classical equilibrium thermodynamics. While this type of behavior is not of a direct interest to hydrodynamics and rheology,  correct mathematical models have to reproduce it. Due to the  universal nature of some of the "hydrodynamically and rheologically uninteresting" observations the requirement that the governing equations reproduce them can  be expected  to provide a universal framework for the governing equations. The requirement of the mass, momentum, and energy conservation does indeed provide a framework for the hydrodynamic equations (the equations have the form of  local conservation laws: time derivative equals divergence of a flux). This  framework can be extended to rheological equations by requiring in addition either the compatibility with the equilibrium thermodynamics \cite{Gr84}, \cite{GrPhys}, \cite{Beris}, \cite{GO}, \cite{OG}, \cite{Grmcontact}, \cite{Grcont}  or the compatibility with the Onsager variational principle \cite{Doi}. In this paper we compare these two types of requirements.

\section{Dynamic and Static Maximum Entropy \\ Principle, GENERIC}\label{Sec1}

In this section we put our attention on   the observed approach of externally (and internally) unforced complex fluids to thermodynamically equilibrium states at which their behaviour is found to be well described by the classical equilibrium thermodynamics. This type of behavior is universally observed. It is the observation that provides  the basis for the classical equilibrium thermodynamics. It was Boltzmann \cite{Boltzmann} who  investigated it first theoretically. His investigation was limited to ideal gases but the mathematical structure of his equation (the Boltzmann kinetic equation) describing the approach to equilibrium  turned out \cite{Gr84}, \cite{GrPhys}, \cite{GO}, \cite{OG} to be applicable to much larger class of macroscopic systems seen on many different scales. The time evolution equations describing the approach to equilibrium are particular realizations of
the abstract Boltzmann equation \cite{Gr84}, called GENERIC \cite{GO}, \cite{OG}
\begin{equation}\label{eqgeneric}
\dot{x}=\frac{1}{\mathbb{E}^*}L\Phi_x(x,\mathbb{E}^*,\mathbb{N}^*)-\Xi_{x^*}(x,x^*)|_{x^*=\Phi_x(x,\mathbb{E}^*,\mathbb{N}^*)}
\end{equation}
By the symbol $x\in M$ we denote in (\ref{eqgeneric}) the variable chosen to describe states of the macroscopic system under investigation, $M$ denotes the state space, $\dot{x}=\frac{dx}{dt}$,  $\Phi:M\times\mathbb{R}^2\rightarrow \mathbb{R}$ is the thermodynamic potential, $L$ is a bivector expressing the kinematics of the state variable $x$, $\Xi:M\times M^*\rightarrow \mathbb{R}$ is the dissipation potential, $\mathbb{E}^*=\frac{1}{T}, \mathbb{N}^*=-\frac{\mu}{T}$,  $T>0$ is an absolute temperature, and $\mu$ is a chemical potential. The space $M^*\ni x^*$ is dual of the space $M$.
We use  the shorthand notation: $A_x(x)=\frac{\partial A(x)}{\partial x}$ where $A$ is a function of $x$  (if $x$ is an element of an infinite dimensional space then $A_x$ is an appropriate functional derivative).

We now explain the physical meaning of the symbols appearing in (\ref{eqgeneric}) and list their properties guaranteeing approach of solutions to (\ref{eqgeneric}) to thermodynamic equilibrium states.
\\

\textit{Thermodynamic potential}
\\

\begin{equation}\label{Phi}
\Phi(x,\mathbb{E}^*,\mathbb{N}^*)=-S(x)+\mathbb{E}^*E(x)+\mathbb{N}^*N(x)
\end{equation}
is required to be a convex function of $x$, $S:M\rightarrow\mathbb{R}$ is entropy, $E:M\rightarrow\mathbb{R}$ is energy, $N:M\rightarrow\mathbb{R}$ is the number of moles, Solutions to
\begin{equation}\label{Phix}
\Phi_x=0
\end{equation}
are thermodynamic equilibrium states $\hat{x}(\mathbb{E}^*,\mathbb{N}^*)$.

State variables in equilibrium thermodynamics are $(\mathbb{E},\mathbb{N}, \mathbb{V})$ denoting  energy,  number of moles and  volume. First, we restrict our analysis to  $\mathbb{V}=1$, we omit $\mathbb{V}$ in the list of state variables,  and interpret  $(\mathbb{E},\mathbb{N})$ as energy and number of moles per unit volume.   We do not change the notation, absence of $\mathbb{V}$ in state variables  means that $\mathbb{E}$ and $\mathbb{N}$ are per unit volume.   The  fundamental equilibrium thermodynamic  relation is
\begin{equation}\label{feq}
\mathbb{S}=\mathbb{S}(\mathbb{E},\mathbb{N})
\end{equation}
where $\mathbb{S}$ is the equilibrium thermodynamic entropy per unit volume.

The fundamental thermodynamic relation on the mesoscopic level on which $x$ serves as the state variable is
\begin{equation}\label{meq}
E=E(x);\,\,N=N(x);\,\,S=S(x)
\end{equation}
The passage from (\ref{meq}) to (\ref{feq}) is made by a sequence of two Legendre transformations. The first one is from (\ref{meq}) to
\begin{equation}\label{fundeq}
\mathbb{S}^*(\mathbb{E}^*,\mathbb{N}^*)=\Phi(\hat{x}(\mathbb{E}^*,\mathbb{N}^*),\mathbb{E}^*,\mathbb{N}^*)
\end{equation}
followed by the second from $\mathbb{S}^*(\mathbb{E}^*,\mathbb{N}^*)$ to
\begin{equation}\label{fundeq1}
\mathbb{S}(\mathbb{E},\mathbb{N})=\widehat{\Phi}(\widehat{\mathbb{E}}^*(\mathbb{E},\mathbb{N}),\mathbb{E},\mathbb{N})
\end{equation}
where
\begin{equation}\label{Phistar}
\widehat{\Phi}^*(\mathbb{E}^*,\mathbb{N}^*,\mathbb{E},\mathbb{N})=-\mathbb{S}^*(\mathbb{E}^*,\mathbb{N}^*)+\mathbb{E}\mathbb{E}^*+\mathbb{N}\mathbb{N}^*
\end{equation}
and $\widehat{\mathbb{E}}^*(\mathbb{E},\mathbb{N})$ is a solution of $\widehat{\Phi}^*_{\mathbb{E}^*}=0;\widehat{\Phi}^*_{\mathbb{N}^*}=0$. It still
remains to extend (\ref{fundeq1}) to
\begin{equation}\label{fundeq2}
\mathbb{S}=\mathbb{S}(\mathbb{E},\mathbb{N},\mathbb{V})
\end{equation}
In equilibrium thermodynamics the fundamental relation (\ref{fundeq2}) is required to be 1-homogeneous,
$\mathbb{S}=\frac{1}{\lambda}\mathbb{S}(\lambda \mathbb{E},\lambda \mathbb{N},\lambda\mathbb{V})$, $\forall \lambda$,  which  expresses mathematically the extensivity of $(\mathbb{S},\mathbb{E},\mathbb{N},\mathbb{V})$. The 1-homogeneity
 then implies (due to the Euler relation) that (\ref{fundeq1}) extends to (\ref{fundeq2}), and $\mathbb{S}^*(T,\mu)=-\frac{PV}{T}$, where $P$ is the conjugate variable to $\mathbb{V}$ having the physical interpretation of the pressure.
 \\

\textit{Bivector $L$}
\\

In the absence of the second term on the right hand side of (\ref{eqgeneric}) the GENERIC equation (\ref{eqgeneric}) represents the Hamiltonian dynamics. This means that $L$ is a Poisson bivector, i.e. the bracket
\begin{equation}\label{PB}
\{A,B\}=<A_x,LB_x>
\end{equation}
is a Poisson bracket,  $<,>$ denotes the scalar product. A bracket $\{A,B\}=<A_x,LB_x>$ is a Poisson bracket if $\{A,B\}=-\{B,A\}$ and the Jacobi identity $\{A,\{B,C\}\}+\{C,\{A,B\}\}+\{B,\{C,A\}\}=0$ holds. The Jacobi identity guarantees that the Poisson bracket $\{A,B\}$ is preserved in the Hamiltonian time evolution $\dot{A}=\{A,E\}$. Indeed, $\dot{\{A,B\}}=\{\dot{A},B\}+\{A,\dot{B}\}=\{\{A,E\},B\}+\{A,\{B,E\}\}$. This equals to $\dot{\{A,B\}}=\{\{A,B\},E\}$  only if  the Jacobi identity holds. In the presence of the second terms on the right hand side of (\ref{eqgeneric}) the Poisson bracket is not conserved even if the  Jacobi identity holds. The Jacobi identity  looses its role in the GENERIC time evolution (\ref{eqgeneric}). For this reason we do not require the Jacobi identity to hold in the GENERIC time evolution.
In the setting of contact geometry in which an appropriately lifted GENERIC (\ref{eqgeneric}) recovers its geometrical clarity \cite{Grmcontact} (the contact structure is preserved in the lifted  time evolution) the Jacobi identity does not play any role.
\\

\textit{Dissipation potential $\Xi$}
\\

The dissipation potential $\Xi(x,x^*)$ is required to satisfy the following properties: (i) $\Xi(x,x^*)|_{x^*=0}=0 \forall x$, (ii) $\Xi(x,x^*)$ reaches its minimum at $x^*=0 \forall x$, (iii) $\Xi(x,x^*)$ is a convex function of $x^*$ in a neighborhood of $x^*=0 \forall x$, (iv) $\Xi(x,x^*)$ depends on $x^*$ only through its dependence on $Kx^*$, called a dissipation potential,  $K$ is linear operator.
We note that the above four requirements imply that in a small neighborhood of $x^*$ all dissipation potentials are quadratic: $\Xi(x,x^*)=<x^*,K^T\Lambda K x^*>$, where $\Lambda$ is a linear positive definite operator.

An additional requirement is the  degeneracy $KE_x=0$ and $KN_x=0$ of the linear operator $K$. This requirement,
together with an additional requirement of degeneracy $LS_x=0$ and $LN_x=0$ of the Poisson bivector $L$,   would guarantee conservation of the energy and the number of moles in the GENERIC time evolution. But such conservation will not hold if a part of the energy (the part  that cannot be expressed in terms of the mesoscopic state variable $x$)  is expressed as a heat (i.e. as a part of the entropy multiplied by the temperature). Similarly in the case of number of moles. We therefore  do not require that the Poisson bivector $L$  and the linear operator $K$ are degenerate.
\\

\textit{Solutions of the GENERIC equation (\ref{eqgeneric})}
\\

With the above specifications,  the GENERIC equation (\ref{eqgeneric}) implies
\begin{equation}\label{SE}
\dot{\Phi}=-<x^*,\Xi_{x^*}>\leq 0
\end{equation}
which gives the thermodynamic potential $\Phi$ the role of the Lyapunov function for the approach of its solutions  to equilibrium states $\hat{x}(\mathbb{E}^*,\mathbb{N}^*)$ that are solutions to (\ref{Phix}).

Summing up,
there are two routes   from the mesoscopic level of description that uses $x$ as the state variable to the level of  equilibrium thermodynamics that uses $(\mathbb{E},\mathbb{N},\mathbb{V})$ as state variables. The first by following the time evolution of the GENERIC equation (\ref{eqgeneric}) and the second by simply finding minimum of the thermodynamic potential $\Phi$. The first is called a dynamic maximum entropy principle,  \textit{dynamic MaxEnt}, the second a static maximum entropy principle, \textit{static MaxEnt}.
\\

\textit{Illustration}
\\

We illustrate an  application  of the GENERIC equation (\ref{eqgeneric}) in rheological modeling of complex fluids on the example of  polymeric fluids. We choose the conformation tensor $\cc=c_{ij}(\rr); i,j =1,2,3$ to characterize their internal structure.  The vector $\rr\in\mathbb{R}^3$ is the position vector, $\cc(\rr)$ is a symmetric and positive definite tensor. We can regard  it as a tensor  expressing deformations of the internal structure. We restrict our analysis in this illustration  to  isothermal fluids so that the complete set of state variables is
\begin{equation}\label{uc}
x=( \rho(\rr),\uu(\rr),\cc(\rr))
\end{equation}
where $\rho(\rr)$ is the mass field and $\uu(\rr)$ is the momentum field.

The first question in the rheological modeling that uses the GENERIC equation (\ref{eqgeneric}) as its framework is of what is the Hamiltonian kinematics of the chosen state variables (\ref{uc}). The polymeric fluids  are seen as a continuum $\mathbb{R}^3$. Transformations $\mathbb{R}^3\rightarrow \mathbb{R}^3$, expressing mathematically  motion of the continuum,   form  a Lie group. From the general theory of Lie groups we know that dual of the Lie algebra is equipped with the Poisson structure. For the Lie group of transformations $\mathbb{R}^3\rightarrow \mathbb{R}^3$ the element of the dual of its Lie algebra is the momentum field $\uu(\rr)$ and the Poisson bracket expressing mathematically its kinematics is
\begin{equation}\label{PB0}
\{A,B\}=\int d\rr \left[u_i(\partial_jA_{u_i}B_{u_j}-\partial_jB_{u_i}A_{u_j})\right]
\end{equation}
where $\partial_i=\frac{\partial}{\partial r_i}$, the symbols $A$ and $B$ denote real valued and sufficiently regular functions of $\uu(\rr)$. We use hereafter the summation convention (summation over repeated indices). The two remaining fields $(\rho(\rr), \cc(\rr))$ in (\ref{uc}) are let to be simply passively advected (Lie dragged) with the flow $\uu$. Again, using general theory of Lie groups, this means that the Poisson bracket expressing kinematics of (\ref{uc}) is
\begin{eqnarray}\label{PBuc}
\{A,B\}&=&\int d\rr \left[u_i(\partial_jA_{u_i}B_{u_j}-\partial_jB_{u_i}A_{u_j})\right] \nonumber \\
&&+\int d\rr \left[\rho(\partial_iA_{\rho}B_{u_i}-\partial_iB_{\rho}A_{u_i})\right]\nonumber \\
&& +\int d\rr \left[c_{ij}(\partial_k A_{c_{ij}}B_{u_k}-\partial_k B_{c_{ij}}A_{u_k})\right.\nonumber \\
&&\left.+c_{ij}((A_{c_{kj}}+A_{c_{jk}})\partial_iB_{u_k}-
((B_{c_{kj}}+B_{c_{jk}})\partial_iA_{u_k}))\right]
\end{eqnarray}
This Poisson bracket has been introduced into rheological modeling in \cite{GrmPB}.
Its detailed derivation can  be found in \cite{book}.
The relation (\ref{PB}) between Poisson bracket and Poisson bivector $L$ provides the Poisson bivector corresponding to the Poisson bracket (\ref{PBuc}).

 .

In the next step in the rheological modeling that uses GENERIC as its framework  we turn our attention from   state variables and their kinematics to the specific nature of the polymeric fluids under investigation. We express it in
the energy $E(\uu,\cc)$, the entropy $S(\uu,\cc)$, and the dissipation potential $\Xi$.  First we discuss $E(\rho,\uu,\cc)$ and $S(\rho, \uu,\cc)$.  Due to our limitation to  isothermal fluids it suffices to specify only the thermodynamic potential (\ref{Phi}). We may choose for instance
\begin{eqnarray}\label{Phiuc}
\Phi(\rho,\uu,\cc, T,\mu)&=&\Phi^{(hyd)}(\rho,\uu,T,\mu)+\Phi^{(int)}(\cc,T)\nonumber \\
\Phi^{(int)}(\rho,\uu,T,\mu)&=&-\int d\rr \left(\frac{1}{2}\ln\det\cc -\frac{1}{T}\epsilon(\cc)\right)
\end{eqnarray}
A detailed derivation of (\ref{Phiuc}) as well as the list of references where this thermodynamic potential was initially derived can  be found in \cite{book}. The first term in (\ref{Phiuc}) is the entropy, $T$ is the constant temperature, $\mu$ is the chemical potential,  $\epsilon(\cc)$ is the internal energy of the internal structure (for instance $\epsilon(\cc)=H tr\cc$ if we model the macromolecules as Hookean elastic dumbbells, $H$ is a constant).

Finally, we have to specify the dissipation potential $\Xi$. We assume that the origin of the dissipation is in the internal structure. We let therefore the dissipation potential $\Xi$ depend only on $(\cc,\cc^*)$. An example of its particular choice is the quadratic dissipation potential  $\Xi(\cc,\cc^*)=\int d\rr \Lambda c_{ij}c*_{jk}c^*_{ik}$ with $\Lambda>0$ (recall that $\cc$ is a positive definite tensor).
From (\ref{eqgeneric}) (in which we replace, due to our limitation to  isothermal fluids, the
two potentials $E$ and $S$  by one potential $\Phi$),  we arrive at the inequality (\ref{SE}).

With the above specifications the GENERIC equation (\ref{eqgeneric})   becomes
\begin{eqnarray}\label{ceqsu}
\frac{\partial\rho}{\partial t}&=&-\partial_j(\rho u_j^*)\nonumber \\
\frac{\partial u_i}{\partial t}&=&-\partial_j(u_iu_j^*)-\partial_ip +\mathfrak{V}_i\nonumber \\
p&=&-\varphi+\rho \rho^*+u_ju^*_j+c_{ij}c^*_{ij}
\end{eqnarray}
\begin{equation}\label{tau}
\mathfrak{V}_i=-\partial_k(c_{kj}(c^*_{ij}+c^*_{ji}))
\end{equation}
and
\begin{equation}\label{ceqsc1}
\frac{\partial c_{ij}}{\partial t}= \mathfrak{F}_i-\Xi_{c_{ij}^*}
\end{equation}
\begin{equation}\label{ceqsc2}
\mathfrak{F}_i=-\partial_k(c_{ij}u_k^*)+c_{kj}\partial_ku_i^*+c_{ki}\partial_ku_j^*
\end{equation}
where $\varphi(\rr)$ is the local thermodynamic potential, i.e. $\Phi=\int d\rr \varphi(\rr)$, and $\uu^*=\Phi_{\uu}; \cc^*=\Phi_{\cc}$.

Now we turn to solutions of Eqs.(\ref{ceqsu}) - (\ref{ceqsc2}). Since these equations are a particular realization of the GENERIC equation (\ref{eqgeneric})   the relations (\ref{SE}) hold and consequently  the thermodynamic potential $\Phi$ plays the role of the Lyapunov function for  the approach to thermodynamic equilibrium states. In other words, solutions to Eqs.(\ref{ceqsu}) - (\ref{ceqsc2}) demonstrate the compatibility with equilibrium thermodynamics.

In addition, we note that  solutions to Eqs.(\ref{ceqsu}) - (\ref{ceqsc2})  also demonstrate the framework of local conservation laws.
Due to our limitation  to isothermal fluids the conserved quantities are the total mass $\int d\rr \rho(\rr)$ and the total momentum $\int d\rr \uu(\rr)$.  We see indeed that the right hand side of the equations governing the time evolution of the mass field $\rho(\rr)$ and the momentum field $\uu(\rr)$ have the form of  divergence of a flux. If its right hand side is integrated $\int_{\Omega}d\rr$  the resulting expression  depends  only on  values on the boundary $\partial\Omega$. It is very important to note that the second  equation in (\ref{ceqsu}) demonstrates even more than the momentum conservation. The expression  for the scalar pressure $p(\rr)$  (that is   implied by the Poisson bracket (\ref{PBuc})) is the formulation of the local equilibrium. This means that the GENERIC formulation of the classical hydrodynamic equations includes both, the mass and momentum conservations and the assumption of local equilibrium. In the classical formulation of the hydrodynamic equations, that is based only on the requirement of the local conservations,  this assumption of local equilibrium is an additional assumption.

There is still another feature of the time evolution of complex fluids that we see in
Eqs.(\ref{ceqsu}) - (\ref{ceqsc2}). The equation governing the time evolution of the internal structure does not have the form of the local conservation law. The internal structure is not conserved during the time evolution. Since in complex fluids the internal structure evolves in time on the same time scale as hydrodynamic fields, the equation governing its time evolution is one of the governing equations. But since it  is not a local conservation law, the extension of hydrodynamics to rheology calls for new frameworks that include the framework for hydrodynamic fields but also address the internal structure. In this paper we discuss two such frameworks: the Onsager principle and GENERIC.

\section{Dynamic and Static Onsager Principle}\label{Sec2}

Equations  (\ref{ceqsu}) - (\ref{ceqsc2}) are a particular realization of dynamic MaxEnt principle. Solutions to (\ref{ceqsu}) - (\ref{ceqsc2}) approach thermodynamic equilibrium states at which equilibrium thermodynamics describes their behavior. This is the way we saw Eqs.(\ref{ceqsu}) - (\ref{ceqsc2}) in the previous section. In this section we see them differently. We see them as the classical fluid mechanics (\ref{ceqsu})  extended by coupling it (by (\ref{tau}), (\ref{ceqsc2})) to an internal structure whose time evolution is governed by (\ref{ceqsc1}). Our focus in this section is put on the approach of the extended GENERIC to the original GENERIC. In the context of Eqs.(\ref{ceqsu}) - (\ref{ceqsc2}) this means that  we solve them in two steps. First, we solve  (\ref{ceqsc1})
 with $\mathfrak{F}$ seen as a fixed  external force. In the second step we insert the solution of (\ref{ceqsc1}) to (\ref{tau}) and solve the GENERIC equation (\ref{ceqsu}) that is the classical hydrodynamic equation adapted to polymeric fluids by the stress tensor (\ref{tau}). We formulate first the two-step view of (\ref{ceqsu}) - (\ref{ceqsc2})  in general terms and then in the context of the polymeric fluids.

 We extend the GENERIC dynamics (\ref{eqgeneric}) to
\begin{equation}\label{eqgenericext}
\left(\begin{array}{cc}\dot{x}\\ \dot{y}\end{array}\right)=\left(\begin{array}{cc}L&K\\-K^T&0\end{array}\right)\left(\begin{array}{cc}x^*
\\y^*\end{array}\right)
-\left(\begin{array}{cc}0\\ \Upsilon_{y^*}(x,y^*)\end{array}\right)
\end{equation}
We call it an \textit{extended GENERIC}.  We explain the meaning of the symbols introduced in (\ref{eqgenericext}). By $x$ we denote the state variable in the original (unextended) GENERIC equation $\dot{x}=Lx^*$, $y$ stands for an \textit{extra state variable} expressing an internal structure.
The linear operator $K$ is  called an \textit{anchor map}, $Kx^*$ is called an \textit{extra vector field $\mathfrak{V}$}, $K^Ty^*$ an \textit{external force $\mathfrak{F}$}. By $\Upsilon(x,y^*)$ we denote dissipation potential satisfying all the properties listed in Section \ref{Sec1}.

From the physical point of view,  the passage from (\ref{eqgeneric}) to (\ref{eqgenericext}) is the passage to a more detailed  viewpoint of macroscopic systems. The internal structure characterized by the extra state variable $y$ takes in (\ref{eqgenericext}) an explicit role in the time evolution.  From the mathematical point of view, we note that the extra state variable $y$ enters dynamics in the vector field driving the time evolution of the original state variables $x$. We are thus extending  (\ref{eqgeneric}) by adding  the vector field $Ky^*$ and introducing a new vector field (the right hand side of the second equation in (\ref{eqgenericext}) driving its time evolution. Regarding the dissipation, by letting to dissipate explicitly only $y$ we assume that the dissipation (expressed in $\Upsilon(x,y^*)$) originates in the internal structure. The thermodynamic potential  $\Phi(x,y,\mathbb{E},\mathbb{N})$  extends the thermodynamic potential $\Phi(x,\mathbb{E},\mathbb{N})$, $x^*=\Phi_x; y^*=\Phi_y$.

Now we turn to solutions of (\ref{eqgenericext}).  We note first that (\ref{eqgenericext}) is a particular realization of the GENERIC structure and thus its solutions approach equilibrium states $(\widehat{x},\widehat{y})$ that are solutions to $\Phi_x(x,y,\mathbb{E},\mathbb{N})=0;\,\Phi_y(x,y,\mathbb{E},\mathbb{N})=0$. Both (\ref{eqgeneric}) and (\ref{eqgenericext}) describe approach to equilibrium. With (\ref{eqgenericext}) we can investigate it  in more details. We assume that the internal structure (i.e. $y$) evolves faster than
$x$ and the approach to equilibrium is made in two stages. The first  is the fast time evolution of $y$ in which $x$ is considered to be a fixed parameter. The final outcome of the first stage is $\widehat{y}(x)$. In the second stage only $x$ evolves with the vector field in which $y$ is replaced by $\widehat{y}(x)$. It is the first stage of the approach (describing an approach of one dynamical theory to another dynamical theory involving less details)  that is the dynamic version of the Onsager variational principle. Similarly as in the time evolution describing approach of a dynamical theory to equilibrium we have dynamic and static MaxEnt, we have in the time evolution describing an approach of one dynamical theory to another dynamical theory involving less details \textit{dynamic and static Onsager principle}. The static Onsager principle is the variational principle introduced in \cite{Rayg}, \cite{OnP}, \cite{OM},  \cite{Prig}, \cite{Gyar}. There are two essential differences between the dynamic MaxEnt and the dynamic Onsager principle. First, the former describes the time evolution in the state space, the latter describes the time evolution in the space of vector fields of a dynamical theory involving less details. Second, the latter does not have to be seen as followed by the second stage.  If considered as an autonomous dynamics (i.e. $y^*$ is seen as an independent state variable, $y^*\neq \Phi_y$)  it  also addresses  externally driven systems that do not approach thermodynamic equilibrium. In this sense  the dynamic Onsager principle is more general than the dynamic MaxEnt. In the rest of this section we briefly formulate the two-stage view of (\ref{eqgenericext}) and then illustrate it on (\ref{ceqsu}) - (\ref{ceqsc2}). The illustration and the comparison of the roles of the balance laws, dynamic MaxEnt principle, and dynamic Onsager principle in modeling of flows of polymeric fluids is the main topic  of this paper.
\\

\textit{General formulation of the dynamic Onsager priciple}
\\

We write the  second equation in (\ref{eqgenericext}) in the form
\begin{equation}\label{Ons}
\dot{y}^*=\mathbb{G}\mathfrak{R}_{y^*}
\end{equation}
where
\begin{equation}\label{Ray}
\mathfrak{R}(y^*,x,\mathfrak{F})=-\Upsilon(y^*,x)+<y^*,\mathfrak{F}>
\end{equation}
is called \textit{Rayleighian}, $\mathfrak{F}$ is an \textit{extra force}, $\mathfrak{F}$. The operator $\mathbb{G}$ is a positive definite operator.

If we consider (\ref{Ons}) in the context of (\ref{eqgenericext}) describing approach to thermodynamic equilibrium  then
\begin{equation}\label{Ons1}
y^*=\Phi_y;\,\,\mathbb{G}=(\Phi_{yy})^{-1};\,\,\mathfrak{F}=-K^Tx^*
\end{equation}
If we consider (\ref{Ons}) as an autonomous dynamical system then $y^*$ is an independent state variable, $\mathbb{G}$  is a positive definite operator and $\mathfrak{F}$ is an external force.
Regarding solutions of (\ref{Ons}), we note that $-\mathfrak{R}$ plays the role of the Lyapunov function for the approach $y^*\rightarrow \widehat{y}^*(x,\mathfrak{F})$,  where $\widehat{y}^*$ is a solution to
\begin{equation}\label{Rayy}
\mathfrak{R}_{y^*}(y^*,x,\mathfrak{F})=0
\end{equation}
By comparing (\ref{Phi}), (\ref{Phix}) with (\ref{Ray}), (\ref{Rayy}) we see clearly that   the dissipation
potential $\Upsilon(y^*,\mathfrak{F})$ plays in the Onsager principle the same role as the entropy $S(x)$ plays in MexEnt. The  role that the temperature $T$ and the chemical potential $\mu$ play in MaxEnt  is played in the Onsager principle by the extra force $\mathfrak{F}$.

The second stage in the approach to thermodynamic equilibrium states
is governed by the first equation $\dot{x}=Lx^*+\mathfrak{V}$ in (\ref{eqgenericext}). In order to continue the approach to thermodynamic equilibrium we thus need to know how does the \textit{extra vector field} $\mathfrak{V}$ depend on $\widehat{y}^*$. We see in (\ref{eqgenericext}) that $\mathfrak{V}=K\widehat{y}^*$. Consequently, the solution to Eq.(\ref{Rayy}) is $\widehat{y}^*= \Upsilon^{\dag}_{(y^*)^{\dag}}((y^*)^{\dag},x,\mathfrak{F})|_{(y^*)^{\dag}=\mathfrak{F}}$, where
$\Upsilon^{\dag}((y^*)^{\dag},x,\mathfrak{F})$ is the Legendre transform of $\Upsilon(y^*,x,\mathfrak{F})$.
The second stage in the time evolution leading to thermodynamic equilibrium sates is thus governed by
\begin{equation}\label{Ons2}
\dot{x}=Lx^*+K\Upsilon^{\dag}_{(y^*)^{\dag}}((y^*)^{\dag},x,\mathfrak{F})|_{(y^*)^{\dag}=\mathfrak{F}}
\end{equation}
By comparing this equation with $(\ref{eqgeneric})$ we see how is the dissipation potential $\Upsilon$ (driving the dissipation of the internal structure $y$ in the first stage)  related to the dissipation potential $\Xi$ driving the dissipation of $x$ in the second stage.
It is also very important to note that the imposed force $\mathfrak{F}$ and the extra vector field $\mathfrak{V}$ are related. Both are generated by the anchor map $K$. The former by $K$ and the latter by its transpose $K^T$.

As we have already mentioned, Eq.(\ref{Ray}) can also be seen as an autonomous dynamics that is not followed by an approach to thermodynamic equilibrium states. The quantity
$\mathfrak{F}$ plays  the role of an external force. In such interpretation, Eq.(\ref{Ray}) is a mathematical formulation of the dynamic Onsager principle.  Its original \cite{OnP} static version is $\mathfrak{R}_{y^*}(y^*,\mathfrak{F})=0$. The extended GENERIC (\ref{eqgenericext}) has been introduced in \cite{Ess1},\cite{Ess2}, \cite{GrRT}.
\\

\textit{Illustration}
\\

We now return to the rheological modeling that we started in the previous section in the illustration. Inspired by (\ref{eqgenericext}), we rewrite (\ref{ceqsu}) into the form
\begin{equation}\label{cRay}
\left(\begin{array}{ccc}\dot{\rho}\\ \dot{u}_i\\ \dot{c}_{ij}\end{array}\right)= L\left(\begin{array}{cc}\rho^*\\ \uu^*\end{array}\right)
+\mathbb{L}^{(I)}\left(\begin{array}{ccc}\rho^*\\ \uu^*\\ \cc^*\end{array} \right) + +\mathbb{L}^{(II)}\left(\begin{array}{ccc}\rho^*\\ \uu^*\\ \cc^*\end{array} \right)
\end{equation}
in which (\ref{ceqsu}) appears as an extension of the classical hydrodynamics with $(\rho(\rr),\uu(\rr))$ playing the role of state variables. The Poisson bivector $L$ appearing in (\ref{cRay}) is $\{A,B\}=<A_x,LB_x>$ where $x=(\rho(\rr),\uu(\rr))$ and
\begin{eqnarray}\label{AB}
\{A,B\}&=&\int d\rr \left[u_i(\partial_j(A_{u_i})B_{u_j}-\partial_j(B_{u_i})A_{u_j})\right.\nonumber \\
&&+\left.\rho(\partial_j (A_{\rho})B_{u_j}-\partial_j(B_{\rho})A_{u_j})\right]
\end{eqnarray}
The anchor map $K$ is
\begin{eqnarray}\label{L1}
\mathbb{L}^{(\alpha)}&=&\left(\begin{array}{ccc}0&0&0\\0&0&K^{(\alpha)}\\0&-(K^{(\alpha)})^T&0\end{array}\right)
\nonumber \\
\alpha&=&I,\,II
\end{eqnarray}
\begin{eqnarray}\label{Ku}
u_i^*=K_{ijk}^{(I)}c_{jk}^*&=&-c_{jk}\partial_ic_{jk}^*\nonumber \\
u_i^*=K_{j}^{(II)}c^*_{ij}&=&2\partial_k(c_{jk}c^*_{ji})
\end{eqnarray}
The first term on the right hand side of (\ref{cRay}) is not interesting from the rheological point of view since
\begin{eqnarray}\label{pc}
K_{ijk}^{(I)}c_{jk}^*(\rr)&=&-\partial_ip^{(int)}(\rr)\nonumber \\
p^{(int)}(\rr)&=&-\varphi^{(int)}(\rr)+c_{ij}c_{ij}^*
\end{eqnarray}
and thus the internal structure influences only the  scalar pressure $p(\rr)$. Changes in $p(\rr)$ are not registered in rheological measurements.
From the second term on the right hand side of (\ref{cRay}) we see that
\begin{eqnarray}\label{FV}
\mathfrak{F}_{ij}&=& c_{kj}\partial_ku^*_i+c_{ki}\partial_ku^*_j   \nonumber \\
\mathfrak{V}_i&=& -\partial_k(c_{kj}(c^*_{ij}+c^*_{ji}))
\end{eqnarray}

Summing up, we have related the Onsager principle  to the GENERIC formulation of mesoscopic dynamics.  The relation represents a  derivation of the Onsager principle and also its extension   by formulating it as a time evolution equation expressing, in the space of vector fields, approach of a mesoscopic dynamics to another mesoscopic dynamics involving less details. A more detailed analysis of the geometrical aspects of GENERIC and the Onsager principle can be found in \cite{Ess1}, \cite{Ess2}, \cite{GrRT}.

\section{Relations among  the balance laws,   Onsager principle, and GENERIC}\label{Sec3}

Rheology investigates reactions of  complex fluids  to imposed forces. Modeling of the reactions proceeds in two stages:
\begin{eqnarray}\label{RM}
&&\textit{imposed external forces},\,\,\mathfrak{F} \longrightarrow \textit{internal-structure dynamics}\nonumber \\
&& \longrightarrow \textit{changes in the experimentally observed flow behavior},\,\, \mathfrak{V}
\end{eqnarray}
 In order to formulate a rheological model we thus need to specify the external force $\mathfrak{F}(y)$, extra vector fields $\mathfrak{V}(y)$,  and \textit{internal-structure dynamics}.
In this section we recall the physical basis of   three frameworks for the  rheological modeling and compare them. The frameworks are: (i) balance laws, (ii) Onsager principle, (iii) GENERIC.
\\

\textit{Balance laws}
\\

Newtonian mechanics entails  conservations of mass, momentum and energy. In continuum mechanics these three conservations  provide the state variables $\rho(\rr),\uu(\rr),e(\rr))$ (local mass, momentum, and energy)  as well as the framework for their time evolution $\frac{\partial x}{\partial \rr}=-\nabla \JJ^{(x)}(\rr)$ (called balance laws or also local conservation laws, $\nabla=(\partial_1,\partial_2,\partial_3);\, x=(\rho(\rr),\uu(\rr),e(\rr)))$. The former because we look for slowly evolving state variables. The local mass, momentum an energy are expected to evolve slowly since globally they do not evolve at all. The latter since we are excluding from our considerations long range interactions among fluid particles.
Specification of the three fluxes $ (\JJ^{(\rho)},\JJ^{(u)},\JJ^{(e)})$ as functions of $\rho(\rr),\uu(\rr),e(\rr))$  is called a \textit{constitutive relation}. If the imposed force is an imposed flow and the complex fluid is isothermal then the  modeling consists of the specification of the mass and momentum fluxes $(\JJ^{(\rho)}(\rr),\JJ^{(u)}(\rr)$. The imposed momentum $\uu(\rr)$ is the external force that acts on the internal structure $y$ by  $\mathfrak{F}(y)$, and the experimentally observed behavior  is the extra stress tensor  $\mathfrak{V}(y)$.
The balance-law framework does not extend to  dynamics of the internal structure since, at least in general, there are no conservation laws in its time evolution. Additional extra frameworks are needed to model the time evolution of $y$  and to specify $\mathfrak{F}(y)$ and $\mathfrak{V}(y)$.  One such framework  is the Kirkwood 2-particle kinetic theory \cite{Kirkwood}, \cite{Kirkwood1}. The Kirkwood theory  has been carried to rheological modeling  in \cite{Bird}.
\\

\textit{Onsager principle}
\\

The Onsager variational principle provides a framework for modeling the \textit{internal-structure dynamics}. Its use in rheological modeling was pioneered by Doi \cite{Doi}. The Onsager  principle addresses directly the graph (\ref{RM}) and puts no restrictions  on the  external forces $\mathfrak{F}$. Its physical basis is the approach of the time evolution driven by dynamics that  involves  an internal structure of fluids to the time evolution driven by dynamics involving only  hydrodynamic fields.
Its application requires a specification of the dissipation potential $\Upsilon(y^*)$, external force $\mathfrak{F}(y)$, and the extra stress tensor $\mathfrak{V}(y)$. Regarding the dissipation potential $\Upsilon(y^*)$, its  choice $\Upsilon(y^*)=<y^*,\Lambda y^*>$, where $\Lambda$ is a positive definite operator is sufficient provided the complex fluids remain  close to thermodynamic equilibrium.
Specification of $\mathfrak{F}(y)$ and $\mathfrak{V}(y)$ requires additional analysis  of interactions between the internal structure and the time evolution of the hydrodynamic fields  $\rho(\rr),\uu(\rr),e(\rr))$. In this respect the rheological modeling guided by the Onsager principle is the same as the rheological modeling guided by balance laws. Specifications of  $\mathfrak{F}(y)$ and $\mathfrak{V}(y)$ lie outside the frameworks.
\\

\textit{GENERIC}
\\

The  extended GENERIC  (\ref{eqgenericext})  does provide a framework for all three nodes $\mathfrak{F}$, $\mathfrak{V}$, \textit{internal-structure dynamics} in the graph (\ref{RM}). It appears that the nodes are related. For instance a change in the external force (for example a slip in the advection \cite{Schow}) has to be accompanied with a corresponding change in the extra stress tensor. The common framework for the complete graph (\ref{RM}) as well as the relation among its nodes is absent in both the balance-law and the Onsager-principle frameworks.
The essential difference between GENERIC and the two previous views of rheological modeling is that GENERIC  begins with an extension. The enlarged system  includes the complex fluid as well as the external  forces. The enlarged system is then considered as externally unforced and its approach to thermodynamic equilibrium (involving both mechanics and thermodynamics) is investigated. Return to the original setting of a complex fluid subjected to external forces is made by reducing the extended setting. Separate investigations of  mechanics of the external forces and the complex fluid that are needed to calculate $\mathfrak{F}$ and $\mathfrak{V}$ in the balance-law and the Onsager-principle frameworks are not needed in the GENERIC framework. Both $\mathfrak{F}$ and $\mathfrak{V}$ have already arisen
 in the investigation of the approach to thermodynamic equilibrium states of the extended system. The extension-followed-by-reduction strategy in rheological modeling has been developed  only for  imposed flows playing the role of  external forces. Limitation in the choice of external forces is a weak point of the GENERIC framework.

Each  of the three frameworks for rheological modeling has its history, tradition, and associated with it feelings and intuition. Our objective was to explore relations among them. The pool of insights needed in modeling is enlarged. In particular, we put into focus the Onsager principle and GENERIC. For instance, if a modeling that follows  the framework of the Onsager principle reaches the stage where an expression for the extra stress tensor is needed,  we suggest to turn to the GENERIC framework. The expression emerging there is directly related to the way the externally imposed flow acts on the internal structure. On the other hand, if a rheological modeling that follows GENERIC needs to deal with external forces that are different from imposed flows,  we suggest to turn to the Onsager-principle framework. More viewpoints merrier is the modeling.
\\
\\

\textit{Acknowledgement}
\\

The exchange of ideas that took place on the Meeting IWNET2025  in Syros, Greece in June 20225 led me to write this paper. In particular, I have benefited from discussions with Antony Beris, Masao Doi, Vlasis Mavrantzas,  Hans Christian Oettinger, and Peter Van. I thank  Vlasis Mavrantzas, Antony Beris, and Kostas Housiadas for organizing the Meeting.

\end{document}